\documentclass[prl,twocolumn,amsmath,amssymb,english,superscriptaddress]{revtex4-1}

\usepackage[dvipdfm]{graphicx}
\usepackage{graphicx,wrapfig}
\usepackage{natbib}
\usepackage{subfigure}
\usepackage{tabularx}
\usepackage{epsfig}
\usepackage{longtable}
\usepackage{amsfonts}
\usepackage{rotating}
\usepackage{subfigure}

\usepackage{hyperref}
\usepackage{babel}

\usepackage{currfile}

\usepackage{color}

\newcommand{\bea}{\begin{eqnarray}}
	\newcommand{\eea}{\end{eqnarray}}

\newcommand{\be}{\begin{equation}}
	\newcommand{\ee}{\end{equation}}



\begin{document}

\title{Edge state preparation in one dimensional lattice by quantum Lyapunov control}
\author{X. L. Zhao$^{1}$,Z. C. Shi$^{1}$, M. Qin}
\affiliation{School of Physics and Optoelectronic Technology, Dalian
University of Technology, Dalian 116024, China\\}	
\author{X.X.Yi}
\affiliation{Center for Quantum Sciences and School of Physics, Northeast Normal University, Changchun 130024, China\\}
\date{\today}
	
\date{\today}
	
\begin{abstract}
Quantum Lyapunov control uses a feedback control methodology to determine control fields which are applied to control quantum systems in an open-loop way. In this work, we adopt two Lyapunov control schemes to prepare an edge state for a fermionic chain consisted of cold atoms loaded in an optical lattice. Such a chain can be described by the Harper model. Corresponding to the two schemes, state distance and state error Lyapunov functions are considered. The results show that both the schemes are effective to prepare the edge state within a wide range of  parameters. We found that the edge state can be prepared with high fidelity even \textbf{if} there are moderate fluctuations in on-site or hopping potentials. Both control schemes can be extended to similar chains (3$m+d$, $d$=2) of different lengths. Since regular amplitude control field is easier to apply in practice, amplitude-modulated control fields are used to replace the unmodulated one to prepare the edge state. Such control approaches provide tools to explore edge states for one dimensional topological materials.
		
\end{abstract}
	
\maketitle
	
	
\section{introduction} \label{introduction}

Topological materials are thought to be the candidates to realize fault-tolerant quantum computation \cite{NP5378} due to their robustness against perturbations. Usually the topological character is indicated by the emergence of edge states for a bulk system. Recently, there have been a great deal of work to explore topological systems related to such states
\cite{PRL109106402,PRL108220401,PRL110075303,PRL110260405,PRL110180403,NP10664,NN6216,PRB90165412,PRL110076401}. In two dimensional systems such as the Bi$_2$Te$_3$ nanoribbon, manipulation of edge state by modulating a gate voltage has been reported in experiments \cite{NN6216}. Similarly, for topological Bi(111) bilayer nanoribbon, through first-principle
simulations, the desirable edge state engineering can be realized by chemical decoration \cite{PRB90165412}. Besides, in one dimensional lattice systems, by coupling the atomic spin states to a laser-induced periodic Zeeman field, a novel scheme is proposed to manipulate the edge state \cite{PRL110076401}. This attracts both theoretical and experimental interests. Thus
manipulation to the edge state is a way to investigate topological systems.

Compared to traditional solid state systems, cold atoms trapped in optical lattices are excellent simulators to investigate various interesting physical topics such as topological insulators \cite{Kane,RMP831057}. One important reason is that such systems with tunable on-site and hopping potentials provide more controllable platforms to investigate quantum signatures of many-body systems. Once such a system possesses a structure of topological material, manipulating the edge state would be feasible in light of quantum control methodology.

Time-dependence control to quantum systems is valid to dynamically realize specific control goals \cite{nature425937,PRL92187902,PRA85022312,42IEEE,Auto411987,PRA80052316,PRL107177204,PLA378699}.
Among them, quantum Lyapunov control has been investigated widely and applied to realize various kinds of objectives \cite{42IEEE,Auto411987,PRA80052316,PRL107177204,PLA378699}. It is used to design an open-loop controller by simulating the evolution for an artificial closed-loop quantum system. Namely it is a `closed-loop design and open-loop apply' control strategy. The dynamics is governed by Schr\"odinger equation. In this control, after determining the suitable control Hamiltonian, the control fields play an important role. They are designed by making the positive Lyapunov function decreases monotonously. In order to design the control fields, in this work, we employ two quantum Lyapunov control schemes: state distance and state error schemes. These two schemes are so intituled since the Lyapunov functions used in them are based on state distance and state error respectively. Note that the control fields would vanish when the system is asymptotically steered to the target state which locates in a set specified by LaSalle' invariant principle \cite{Lasalle}.

We prepare the edge state for a topological system composed of atoms loaded in an optical lattice by the quantum Lyapunov control methods mentioned above. Such a lattice can be created by two standing waves formed by two laser beams with different wavelengths \cite{pra75061603,nature453895}. Such a system can be described by a Hamiltonian of the Harper model \cite{Harper}. We assume all the cold atoms are loaded in the lowest band of the optical lattice to make the tight-binding limit available. For the open boundary lattice, the edge states emergence with eigenenergies locate apart from the energy subbands. We choose one of the edge states to prepare. Then a control Hamiltonian is needed. Inspired by lattice shaking technique \cite{PRL95260404,PRL95170404,NATURE9769} which can be applied to quantum simulators in optical lattices with tunable structures, in this work, a trigonometrical modulation control Hamiltonian (time-dependent Hermitian) is introduced with a time varying control field to prepare the edge state.

In order to indicate the validity of both control schemes, the fidelity defined by the scalar product of the controlled and goal state is adopted. We found that despite the existence of fluctuations which perturb the on-site or the hopping potentials, the fidelity can reach a high value at the terminated time. This demonstrates the robustness of both control schemes against fluctuations. To examine the generality of both control schemes, we apply them to chains with different lengths with the same configuration(all of them are of type 3$m+d$, $d$=2 \cite{Bohm}). High fidelity control results manifest the validity of both control schemes in this work. Taking the controllability for operation in experiments into account, we use amplitude-modulated control fields in place of the unmodulated ones to prepare the edge state. Since the sign for the control field would be inclined to switch while its amplitude changes when fidelity approaches 1, we choose the manipulation function to make the amplitude decrease.

This paper is organized as follows. In Section \ref{model}, we specify the model, i.e. a controlled chain made of cold atoms loaded in an one dimensional optical lattice. In Section \ref{app}, we exhibit the general procedure to design the control fields in two quantum Lyapunov control schemes, which is exemplified with a trigonometrical modulation control Hamiltonian. Then we examine the control strategies in large control parameter intervals and with random initial states. In Section \ref{discussions}, we explore the robustness of the two control schemes against the number and amplitudes of on-site energy and hopping fluctuations, application to chains of 3$m+d$ configuration with different
lengths and a feasible modulation for the control fields. Finally, we conclude in Section \ref{sum}.

\section{fermionic chain made of atoms loaded in optical lattice} \label{model}
In this work we consider a system made of cold atoms trapped in an one dimensional optical lattice. The lattice of straight line shape can be generated by superimposing two standing waves of laser beams with different wavelengths. First we show the procedure to obtain the fermionic chain from single particle Hamiltonian, which has the following form in the periodically modulated lattice
\begin{eqnarray}  \label{Hs}
\hat{H}_s&=&\hat{H}_1+\hat{H}_2,  \nonumber \\
\hat{H}_1&=&\frac{p_x^2}{2M}+V_1\sin^2(k_1x),  \nonumber \\
\hat{H}_2&=&V_2\cos^2(k_2x+\delta),
\end{eqnarray}
where $V_j=s_jE_{rj}$ and $k_j=2\pi/\lambda_j(j=1,2)$ are the lattice depth and wavenumbers respectively. $x$ denotes the positions for the atoms on the chain. $s_j$ and $E_{rj}=h^2/(2M\lambda_j^2)$ denote the height of the lattices and recoil energies respectively. $h$ is the Planck constant and $M$ is the mass of the atoms in the lattice. $\delta$ is an arbitrary phase of the second laser beam. We assume all the
atoms are trapped in the lowest band of the optical lattice to make the tight-binding approximation available.  Then in virtue of the field operators $\Phi(x)$, the Hamiltonian reads
\begin{eqnarray}
\hat{H}_0=\int dx\Phi^{\dagger}(x)\hat{H}_s\Phi(x).\label{Hs2}
\end{eqnarray}
In the basis of Wannier functions, the field operator can be expanded as $\Phi(x)=\sum_n\hat{c}_n\omega(x-x_n)$. $\hat{c}_n$ here denotes the annihilation operator for the fermion at site $n$ while the spin freedom is not considered. Substitute this into (\ref{Hs2}), omitting constant terms, one gets the Hamiltonian
\begin{eqnarray}
\begin{aligned}
\hat{H}_0=&-J\sum_{n=1} (\hat{c}^\dagger_n \hat{c}_{n+1}+\mathrm{H.c.})\\
&+\sum_{n=1} V\cos(2\pi\beta n+\delta) \hat{n}_n,\label{H00}
\end{aligned}
\end{eqnarray}
where $\beta$=$k_2/k_1$=$p/q$ ($p$ and $q$ are prime to each other), the hopping amplitude $J$=$\int dx \omega^*(x-x_{n})\hat{H}_1\omega(x-x_{n+1})$ and on-site energy $V=\frac{V_2}{2}\int dx \omega^*(x)\cos(2k_2x)\omega(x)$. $\hat{c}^\dagger_n$ ($\hat{c}_n$) are the creation (annihilation) operators for the atoms on-site $n$, and $\hat{n}_n=\hat{c}^\dagger_n \hat{c}_n$.
$J$ is set to be the unit of energy in this work and we set $\hbar$=1. We choose $\beta=1/3$, then the chain has a 3$m+d$ configuration: $m$ here means the number of eigenenergies in a energy subband since the total energy band can be divided into $q$ subbands in Harper model \cite{PRL109106402,Bohm} and we choose $d$=2 in this work. `3' reflects the periodic character for the chain which results from the ratio of the two wave vectors of the two lasers. `$d$' here is the remainder for the length of the chain divided by `3' which is a character for the destruction of the translation symmetry of the chain. To get $J$ and $V$ mentioned above, estimations have been obtained by calculating the integrals in terms of maximally localized Wannier functions \cite{PRA72053606}. They are $J\simeq1.43s_1^{0.98}e^{(-2.07\sqrt{s_1})}E_{r1}$ and $V\simeq\frac{s_2\beta^2}{2}e^{-\frac{\beta^\alpha}{s_1^{\gamma}}}E_{r1}$, where $\alpha$ and $\gamma$ are determined by fitting the numerical evaluation of the integral of $V$. Roughly, Gaussian approximation for the Wannier function can also be used to estimate $J$ and $V$ \cite{njp11033023}. According to these expressions for $J$ and $V$, one can see that several parameters can be adjusted to yield various ratio of $V/J$. Since we focus on the control procedure, moderate value of $V/J$ is chosen directly in this work.
\begin{figure}
	\includegraphics[width=7.7cm]{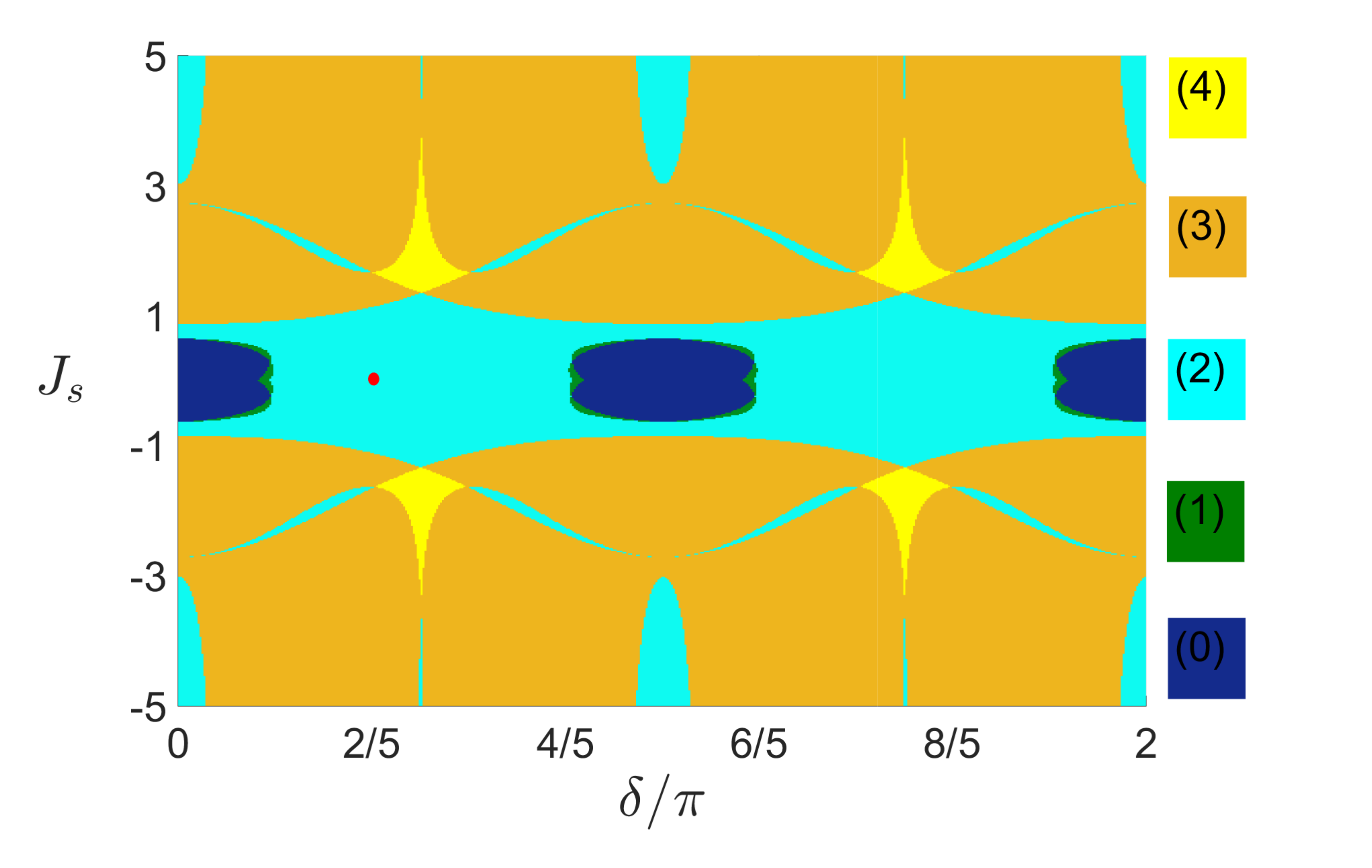}
	\caption{The number of isolated states vs. $J_s$  and $\delta$.
		Different colors show the different total number of isolated states in
		the total energy band, as labeled in the right side. In this work,
		we set the parameters $L=62$, $V=1.5J$, $J_s$=0, and $\delta=2\pi/5$
		(i.e. the red point in this figure) to exhibit the control process.}
	\label{deVp}
\end{figure}
In order to determine the controlled chain, now we numerically find out the spectrum for the eigenenergies as a function of the parameters in the Hamiltonian. The edge states for the chain locate near the ends. The hopping strength between the two ends (denoted by $J_s$) would affect the energy spectrum of the chain while the chain can be mapped to a ring mathematically without changing the periodic character in the bulk. Namely different strength of $J_s$ induce different number of isolated states in the total band. In this work, the eigenenergies corresponding to the isolated states locate apart from the three energy subbands (since we have set $\beta=1/3$). Here, in order to define such isolated states, we assume the energy difference between the isolated one and their nearest neighbors is $E_d$ and the maximum energy difference between the neighbor eigenenergies in the corresponding neighbor subband is $E_b$. Here the `corresponding neighbor subband' refers to the one that the nearest neighbor eigenenergy belongs to. If $E_d>\chi E_b$, $\chi=2$, we refer it to be the isolated state in this work. With respect to mentioned above, in Fig. \ref{deVp}, we numerically explore the number of isolated eigenenergies in the total energy band vs. the distinct hopping strength $J_s$ and the phase $\delta$. It can be seen that there would be more than 2 isolated states appearing in some combinations of $J_s$ and $\delta$. The isolated states include the edge states belonging to topological phases. Besides the edge states, the other isolated ones are just located in the gap since the eigenstates are orthogonal to each other. Numerical simulation shows that the eigenenergies corresponding to the edge states do not always locates in the gap between the subbands but may be larger or smaller than all the other eigenenergies. Here $E_d>2E_b$ is obvious a rough criterion to ascertain an isolated state since the energy difference between the isolated one and its neighbor vary gradually with respect to $J_s$ and $\delta$. So further investigations may be needed on this $\text{rough}$ spectrum.

\begin{figure}
	\includegraphics[width=7.7cm]{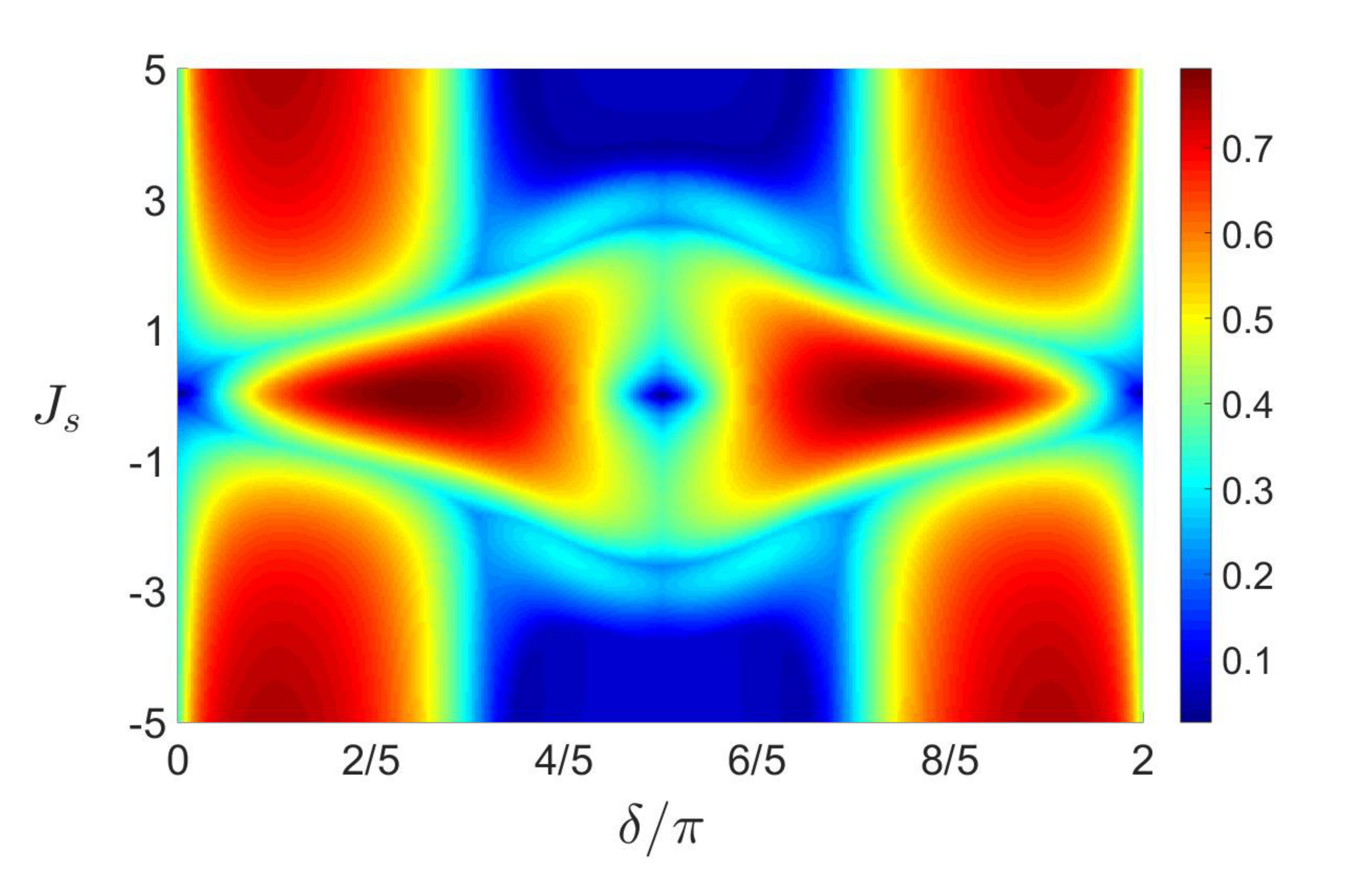}
	\caption{The quantity $I_{m+1}$ to measure the localization for the
		$m$+1 edge states vs. $J_s$ and $\delta$. The other parameters are
		the same as those in Fig. \ref{deVp}.} \label{maxlo}
\end{figure}

As mentioned above, if the criterion for an isolated state $E_d>\chi E_b$ changes, one obtains a different spectrum from Fig. \ref{deVp}. To further determine a model chain, we next examine the character of an edge state which is chosen as the target. Numerical simulations show that the edge state may appear as the $m+1$ eigenstate in the energy band with respect to $J_s$ and $\delta$. Here the eigenstates have been arrayed according to their corresponding eigenenergies in a small-value to large-value manner. Thus $m+1$ refers to the order for the eigenstate in the array. Since a more local edge state is desirable, we next check the localization for the $m+1$ eigenstate. IPR can be used to indicate the degree of localization for a state \cite{PRA76042333}. If a state $|\phi_j\rangle=\sum_{n=1}^N\psi_j(n)|n\rangle$ ($N$ is the number of basis $|n\rangle$), IPR can be defined as $I_j=\sum_{n=1}^N|\psi_j(n)|^4$ ($j$=$m$+1 in this work). We can see that if $\psi_j(n)$ are distributed homogeneously over all basis $|n\rangle$, namely $|\psi_j(n)|^2\sim1/N$, then $I_j\sim1/N$. Whereas, if $\psi_j(n)$ are localized over a range $\zeta$, namely $|\psi_j(n)|^2\sim1/\zeta$, then $I_j\sim1/\zeta$. So for large $N$, the larger $I_j$ is, the more degree of localization of state $|\phi_j\rangle$. Then we show $I_{m+1}$ vs. $J_s$ and $\delta$ in Fig. \ref{maxlo} to pick a combination of $J_s$ and $\delta$ with high IPR. According to this figure, without loss of generality, we choose $J_s=0$ and $\delta=2\pi/5$ mainly to exhibit the control results. According to the mentioned above, we specify the $m+1$ eigenstate as the target in this work. Then the ring retrogresses to an one dimensional chain.

\begin{figure}
	\includegraphics[width=7.7cm]{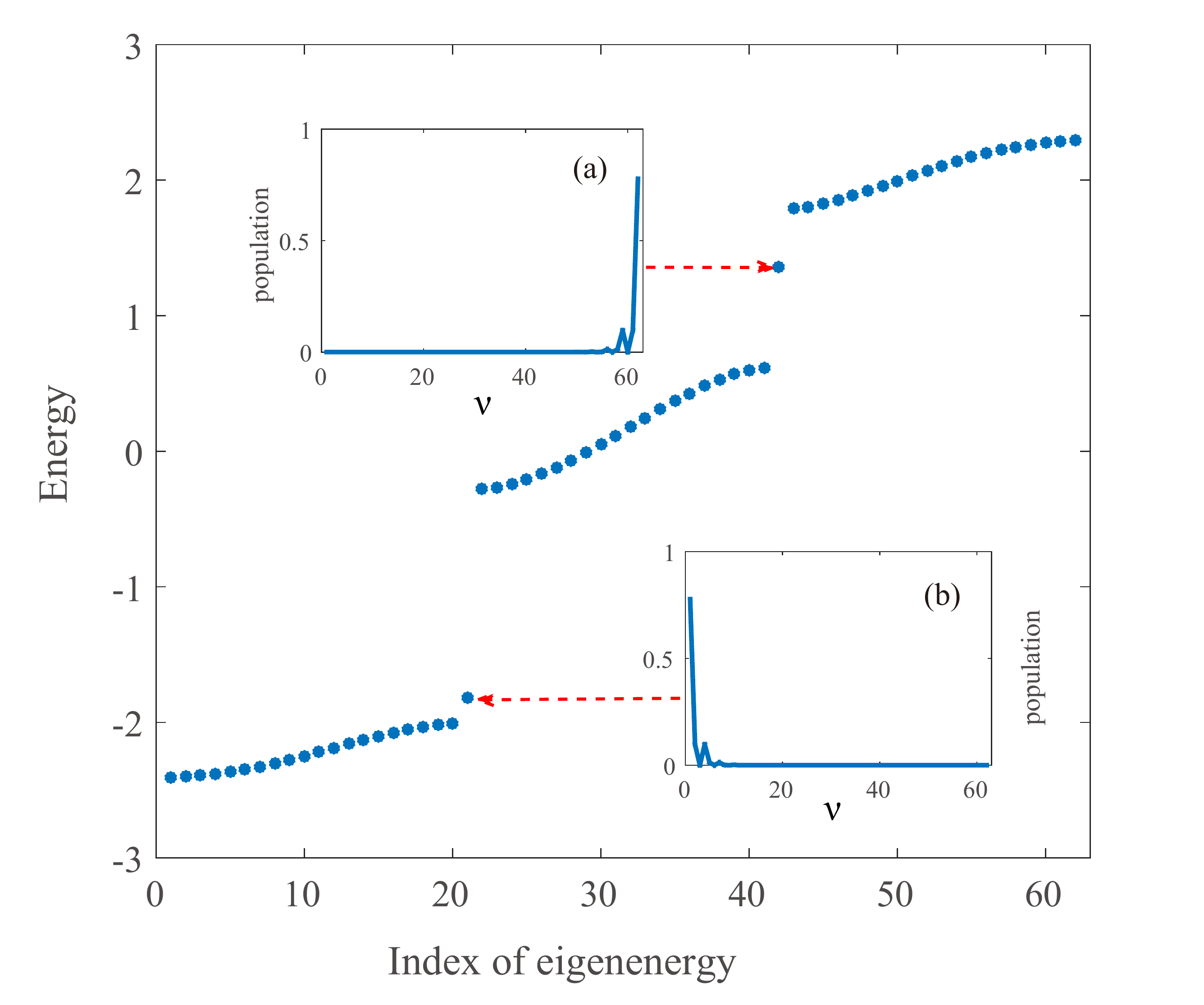}
	\caption{Energy spectrum and the distribution of the edge states for the chain when $J_s$=0, $\delta$=
		$2\pi/5$ as the red point marks in Fig. \ref{deVp}. The edge state as (b) shows is chosen as the
		target state in this work. $\nu$ in the insets are the index for the sites on the chain.} \label{ec}
\end{figure}

Since the model chain has been specified, we next check the energy spectrum for it. This gives us an intuitive knowledge for the target state. Fig. \ref{ec} shows the single-excitation energy spectrum while insets (a) and (b) exhibit the population of the edge state-2($m$+1) and $m$+1-state of Hamiltonian (\ref{H00}) when $\beta$=1/3 in the open boundary condition. It can be intuitively seen that the population of the edge states both localize at the ends. They are the eigenstates of the natural Hamiltonian (\ref{H00}) with the protected eigenenergies \cite{PRL108220401}. We next show the procedure to prepare the `$m+1$' edge state by two kinds of Lyapunov control schemes.

\section{prepare the edge state by Lyapunov control} \label{app}

First we briefly exhibit the design procedure in quantum Lyapunov control. Generally, the controlled system is governed by Schr\"odinger equation
\begin{eqnarray}
i\dot{|\varphi\rangle}=(\hat{H}_0+\sum_{n}f_n(t)\hat{H}_n(t))|\varphi\rangle,\label{dv}
\end{eqnarray}
where $\hat{H}_0$ describes the natural Hamiltonian and $\hat{H}_n(t)$ are the control Hamiltonians with the corresponding real-valued $f_n(t)$. $f_n(t)$ represent the control fields need to be designed by quantum Lyapunov control method. The system state is $|\psi\rangle=\sum_{k=1}^L \xi_k|k\rangle$, where $|k\rangle$ denotes that the atom on site $k$ is excited while the others not, and $\xi_k$ is the corresponding probability amplitude. $L$ denotes the total number of sites on the chain. The control Hamiltonian $\hat{H}_n(t)$ should not commute with the natural Hamiltonian $\hat{H}_0$, (i.e. $[\hat{H}_0,\hat{H}_n(t)]\neq0$), otherwise its effect can be included in the natural Hamiltonian. Note that the target state $|\varphi_f\rangle$ is usually an eigenstate of the natural Hamiltonian, namely $\hat{H}_{0}|\varphi_f\rangle=\lambda_{f}|\varphi_f\rangle$. Then the Lyapunov function $V_L$ related to the controlled state is constructed to design the control field. Then the control fields are determined by making the first-order time derivative of the Lyapunov function $V_L$ negative. Then assisted by the control Hamiltonians with the corresponding designed control fields, the system would be steered to a LaSalle invariant set asymptotically in which the states satisfy $\dot{V}_{L}=0$.

Usually, there are alternatives of Lyapunov functions that can be used. One of the candidates is based on Hilbert-Schmidt distance between the system state $|\varphi(t)\rangle$ and the target state $|\varphi_f\rangle$ \cite{PFIS} (We call it Lyapunov-A for short hereafter). It is
\begin{eqnarray}
V_{A}=\frac{1}{2}(1-|\langle \varphi_f|\varphi(t)\rangle|^{2}),\label{V1}
\end{eqnarray}
where $|\langle \varphi_f|\varphi(t)\rangle|^{2}$ denotes the transition probability from $|\varphi(t)\rangle$ to $|\varphi_f\rangle$. According to the description above, the first-order time derivative for $V_{A}$ is
\begin{eqnarray}
\begin{aligned}
\dot{V}_A=&-\sum_{n}f_{An}(t)\cdot|\langle\varphi(t)|\varphi_f\rangle|\times\\
& Im[e^{i\arg\langle\varphi(t)|\varphi_f\rangle}\langle\varphi_f|\hat{H}_{n}|\varphi(t)\rangle],
\label{Vdot}
\end{aligned}
\end{eqnarray}
where $Im[\bullet]$ denotes the imaginary part of $\bullet$. Thus there are different kinds of control fields $f_{An}(t)$ that meet the requirement $\dot{V}_{A}\leq0$. For example, a succinct and valid choice is
\begin{eqnarray}
\begin{aligned}
f_{An}(t)=T_{n} Im[e^{i\arg\langle\varphi(t)|\varphi_f\rangle}\langle\varphi_f|\hat{H}_{n}|\varphi(t)\rangle],
\end{aligned}\label{fd}
\end{eqnarray}
where $T_{n}>0$. When $\langle\varphi(t)|\varphi_f\rangle=0$, the angle $\arg\langle\varphi(t)|\varphi_f\rangle$ is uncertain. Without loss of generality, we artificially set $\arg\langle\varphi(t)|\varphi_f\rangle=0$ in this situation.

Another Lyapunov function based on state error \cite{Auto411987} can be described by (We call it Lyapunov-B for short hereafter)
\begin{eqnarray}
\begin{aligned}
V_{B}&=\frac{1}{2}\langle\varphi(t)-\varphi_f|\varphi(t)-\varphi_f\rangle\\
&=1-Re[\langle\varphi_f|\varphi(t)\rangle].\label{V2}
\end{aligned}
\end{eqnarray}
$Re[\bullet]$ denotes the real part of $\bullet$. The first-order time derivative for $V_{B}$ is
\begin{eqnarray}
\begin{aligned}
\dot{V}_{B}=&-\lambda_f Im[\langle\varphi_f|\varphi(t)\rangle]\\
&-\sum_nf_{Bn}(t)\cdot Im[\langle\varphi_f|\hat{H}_{n}|\varphi(t)\rangle].\label{dotV2}
\end{aligned}
\end{eqnarray}
Distinguishing from the first Lyapunov function, here we would employ a simple $\hat{H}_{c0}=I$ where $I$ is the identity matrix
in Hilbert space and $f_{B0}=-\lambda_f$ to cancel the first term in (\ref{dotV2}). Therefore, we can choose
\begin{eqnarray}
\begin{aligned}
f_{Bn}(t)=& T_{n} Im[\langle\varphi_f|\hat{H}_{n}|\varphi(t)\rangle],\\
& n\neq0,\label{fr}
f_{B0}=-\lambda_f,
\end{aligned}
\end{eqnarray}
where $T_n>0$. In the next section, we apply both control schemes to generate edge state for the fermionic chain.

To apply Lyapunov control, we next specify the control Hamiltonian. There may be various control Hamiltonians that can be adopted to prepare the edge state for the chain. Taking available techniques into consideration, a trigonometrical shaking Hamiltonian which is generated by electro-optic phase modulator \cite{PRL95170404} is employed, i.e.
\begin{eqnarray}
\hat{H}_r=V_c\cos^2[k_d(x-b\cos(\omega_{c}t))].\label{sk}
\end{eqnarray}
$V_c$ is the constant amplitude and $k_d$ denotes the laser wave vector. $\omega_c$ reflects the shaking frequency. $b$ indicates the shaking depth and should not be large otherwise it may induce heating effect leading to failure of this model. In the second quantization form, the control Hamiltonian reads
\begin{eqnarray}
\hat{H}_c=\int dx \Phi^\dag(x)\hat{H}_r\Phi(x).\label{sk2}
\end{eqnarray}
As all atoms are loaded in the lowest band of the optical lattice, in terms of Wannier basis similar to (\ref{Hs2}), the control Hamiltonian has the matrix element
\begin{eqnarray}
\begin{aligned}
&\hat{H}_c(m,n)\\
&=\delta_{m,n}V_{cd}\cos[2\pi \frac{k_d}{k_1}m+2k_db\cos(\omega_{c}t)],\label{skm}
\end{aligned}
\end{eqnarray}
where $V_{cd}=\frac{V_c}{2}\int dx\omega^*(x)\cos(2k_dx)\omega(x)$ and $m, n$ are integers indicating the order for atoms on the chain.
The parameters $2k_db$ and $\omega_c$ can be modulated in experiments \cite{PRL95170404}. The total Hamiltonian reads $\hat{H}=\hat{H_0}+f_c(t)\hat{H}_c$. Here $f_{c}(t)$ is the time varying control field determined by quantum Lyapunov method. They can be tuned by changing the voltage on the electro-optic phase modulator. In this work, we denote the control fields as $f_A(t)$ in the Lyapunov-A and $f_B(t)$ in the Lyapunov-B schemes respectively.

\begin{figure}
	\includegraphics[width=7.7cm]{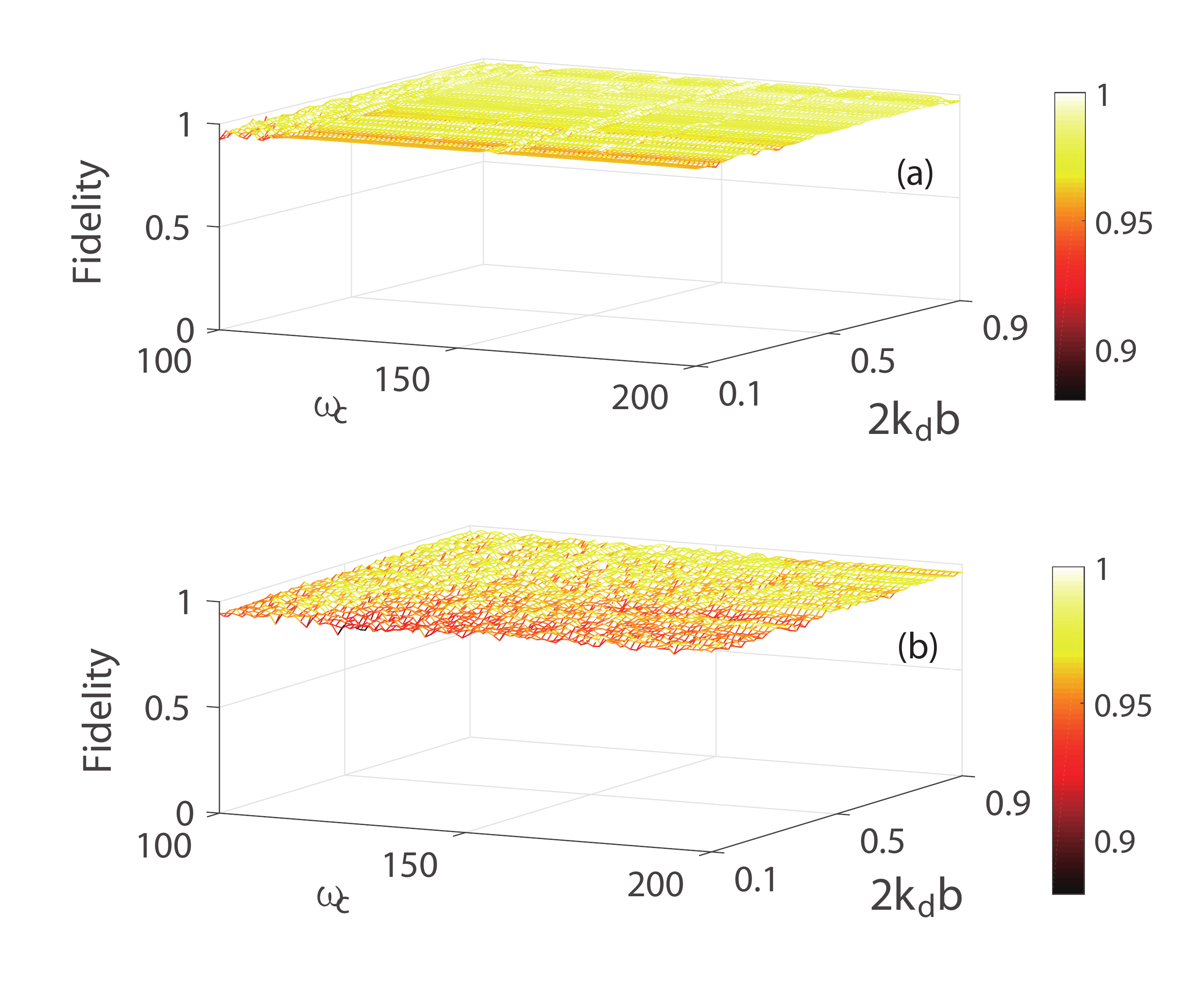}
	\caption{Fidelity at time $t$=3000 (in units of $1/J$) vs.  $2k_db$
		and $\omega_c$ (in units of $J$) for Lyapunov-A and Lyapunov-B
		schemes in (a) and (b) respectively. We have set $V=1.5J$,
		$\beta$=1/3, $\delta=2\pi/5$ in the natural Hamiltonian.
		$V_{cd}$=5$J$ in the control Hamiltonian (\ref{skm}). We have used
		the laser for the control 2$\pi k_d/k_1\simeq$4.97, the chain of
		length $L=62$.} \label{DEOPT}
\end{figure}

\begin{figure}
	\includegraphics*[width=7.7cm]{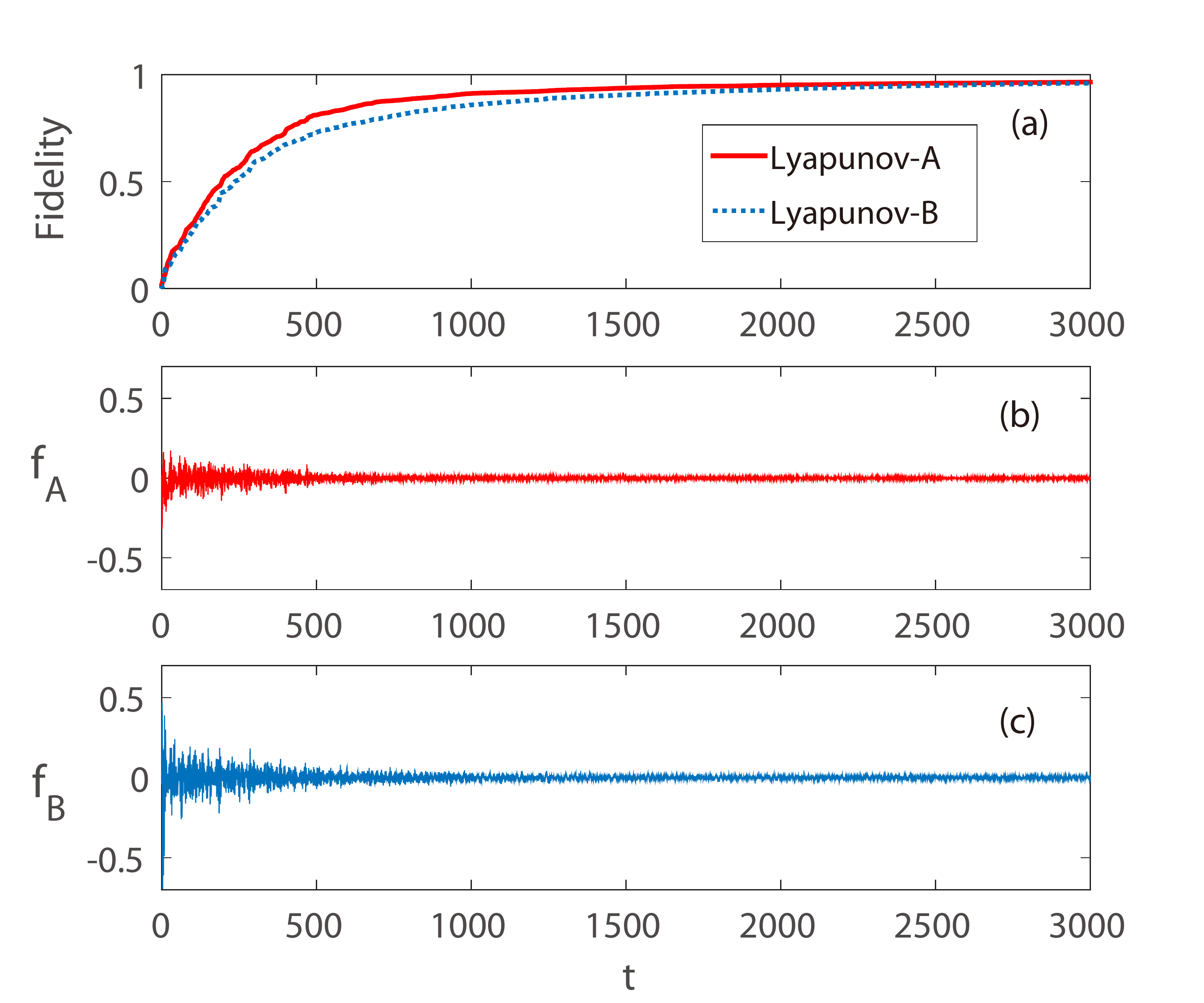}
	\caption{Evolution of the fidelity for the Lyapunov-A and
		Lyapunov-B schemes and the control fields. We have chosen
		$2k_db$=0.2, $\omega_c=115J$ which is available in experimental
		technology \cite{PRL95170404,NATURE9769}. $f_A$  and $f_B$ are the
		control fields for Lyapunov-A and Lyapunov-B schemes respectively.
		The other parameters in natural Hamiltonian are same to those in
		Fig. \ref{DEOPT}.} \label{FDE}
\end{figure}

To exhibit the control effect, we quantify it by the fidelity which is defined as
\begin{eqnarray}
\mathcal{F}(t)=|\langle\varphi(t)|\varphi_{edge}\rangle|^{2},\label{Fid}
\end{eqnarray}
where $|\varphi_{edge}\rangle$ is the target state as is shown by the inset (b) in Fig. \ref{ec}. We examine the fidelity at terminated time $t=3000$ vs. $2k_db$ and $\omega_c$  numerically in Fig. \ref{DEOPT} for the two kinds of control schemes. In these simulations, we have set the initial state with equally projection to the basis $|k\rangle$. It can be seen that both schemes can be used to yield high fidelity in a wide range of parameter interval.

\begin{figure}
	\includegraphics[width=7.7cm]{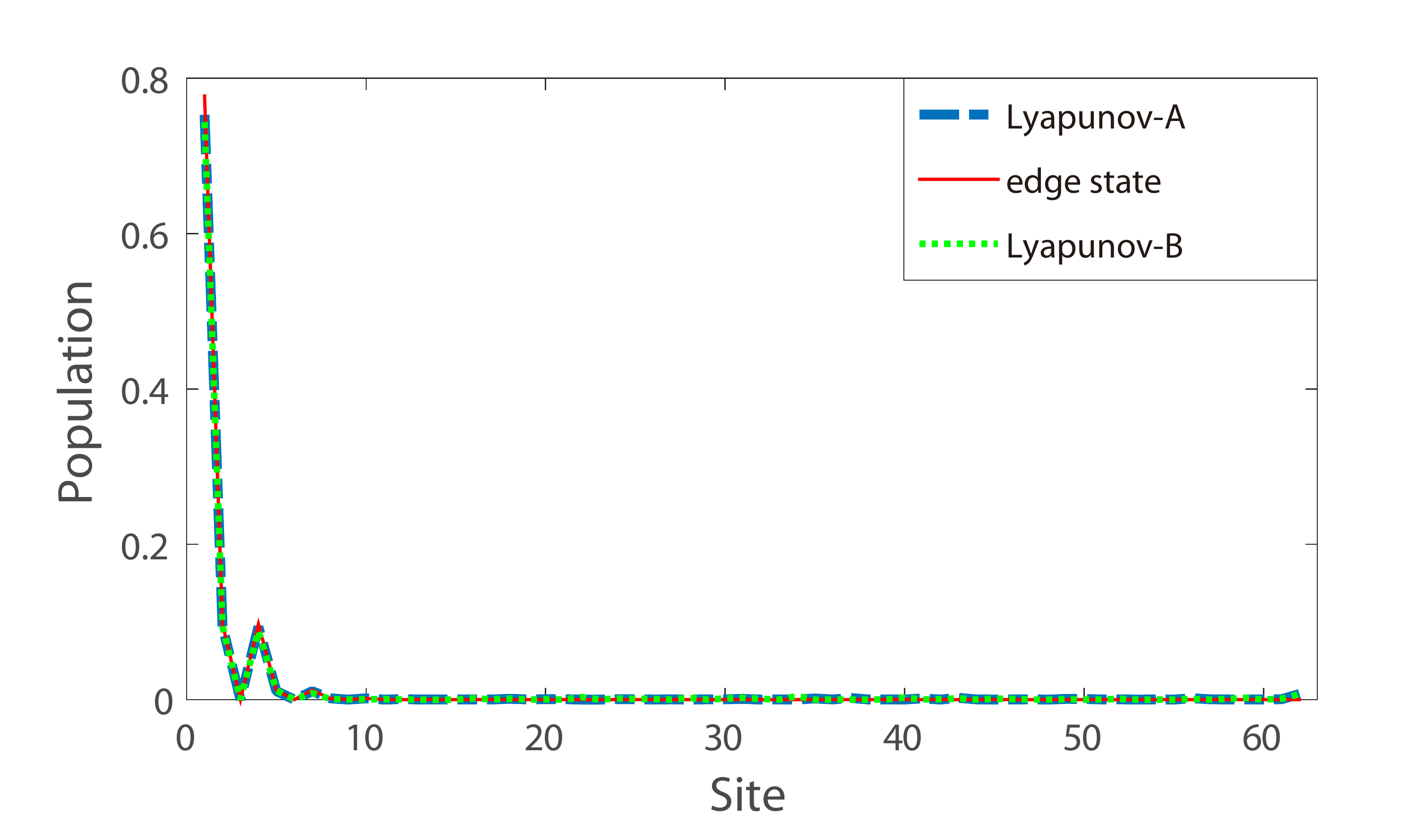}
	\caption{The result density distribution at time $t$=3000  as in
		Fig. \ref{FDE} for the two control schemes and the goal match well.}
	\label{pop}
\end{figure}

To be more concrete, we show the dynamics of the control process in Fig. \ref{FDE} for both the control schemes, when $2k_db$=0.2 and $\omega_c=115J$. Numerical simulation shows that the fidelity can reach more than 0.95 at time $t$=3000 while each control field approximately vanishes. To intuitively compare the control results with the target, in Fig. \ref{pop}, we plot the density distribution at time $t=3000$. It can be seen that the two profiles matches well at the terminated time.

\begin{figure}
	\includegraphics[width=7.7cm]{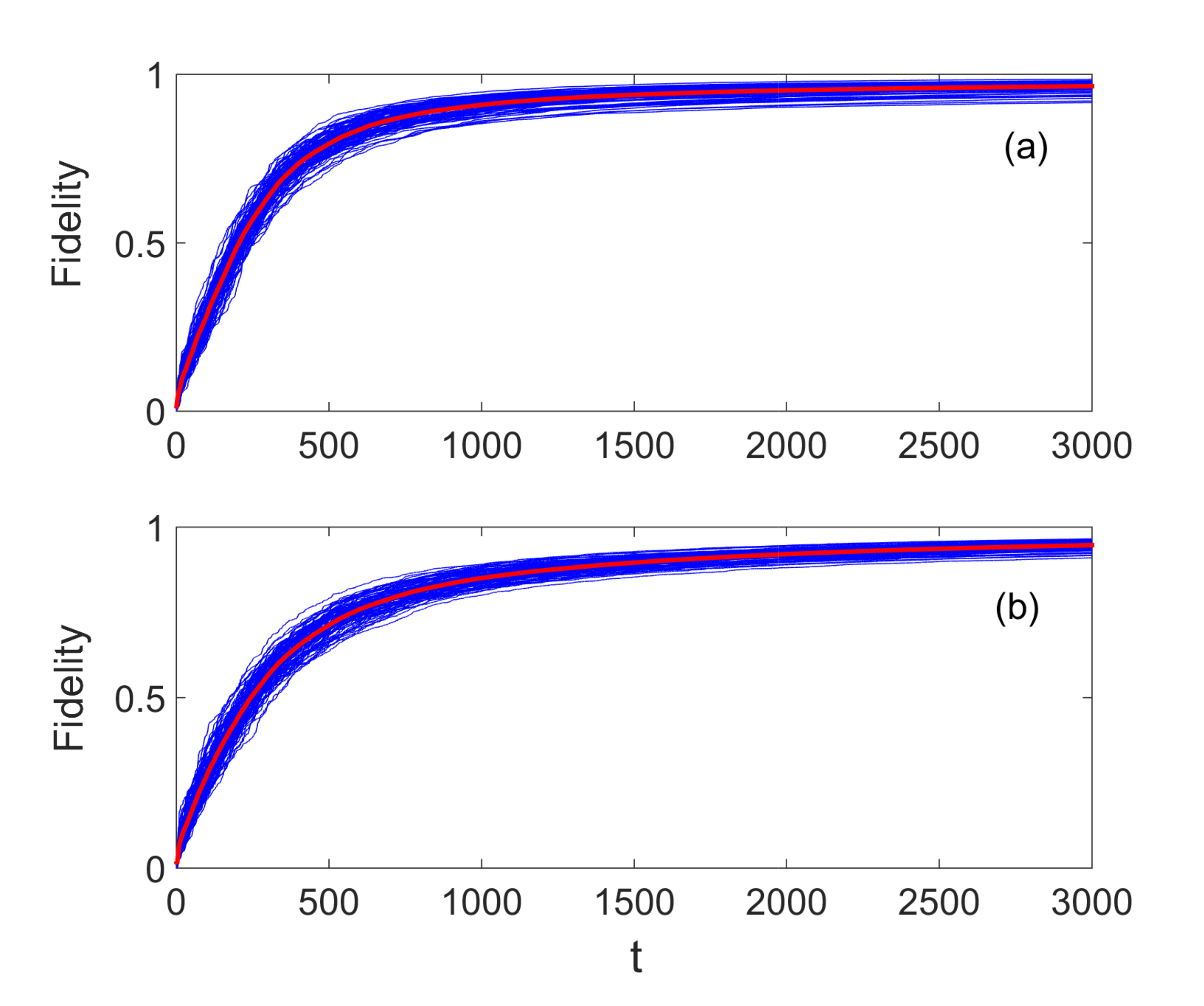}
	\caption{The evolution of the fidelity for 100 random initial
		states in each control scheme. (a) is for the Lyapunov-A scheme
		while (b) is for Lyapunov-B scheme. The red solid lines represent
		the average for the blues.}\label{randDCER}
\end{figure}

In practice there may be various kinds of initial states. In consideration of this, we tested 100 site-occupation random initial states for both kinds of control schemes to check their validity for preparing the edge state in Fig. \ref{randDCER}. It can be seen that both control schemes are effective to complete the control goal.

\section{Discussions} \label{discussions}

\begin{figure}
	\includegraphics[width=8cm]{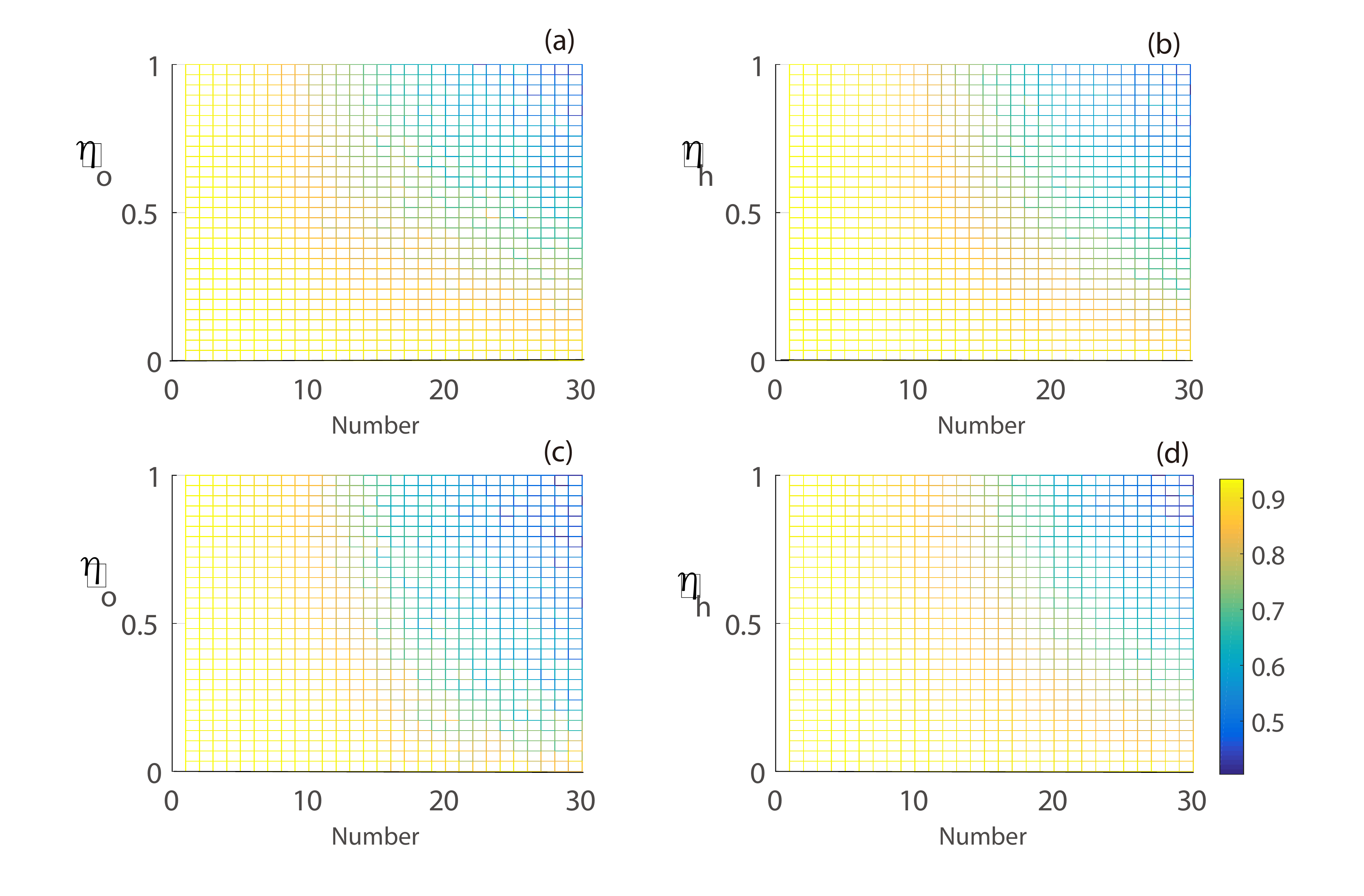}
	\caption{Fidelity at time $t$=3000 vs. the number and the  maximal
		strength of the random fluctuations averaged over 30 times for each
		value. (a) and (b): fidelity for on-site energy and hopping fluctuations
		in Lyapunov-A scheme; (c) and (d): fidelity for on-site energy and
		hopping fluctuations in Lyapunov-B scheme. The parameters are same to
		those in Fig. \ref{FDE} where $\eta_o$ and $\eta_h$ are in units of
		$J$. All the figures have the same color map.} \label{SDEOH}
\end{figure}

In this section we provide discussions for both control schemes with respect to their robustness against fluctuations on the chain, expandability to chains of different lengths and modulation for the control fields.

Inevitably there exist types of fluctuations on the chain which mainly be classified as on-site energy and hopping types. These two kinds of fluctuations can be denoted as $\delta \hat{V}$=$\eta_o\hat{H}_{po}$ and $\delta \hat{J}$=$\eta_h\hat{H}_{ph}$. In matrix form, $\hat{H}_{po}(m,k)=\delta_{m,k}$ and $\hat{H}_{ph}(m,l)=\hat{H}^*_{ph} (l,m)=\delta_{m,l+1}$, here $m, k\in [1,L]$ and $l\in [1,L-1]$ are random integers representing positions for the fluctuations. In the simulation, we set the strength of the fluctuations randomly distributed in the interval: [0,$\eta_o$($\eta_h$)]. Namely, there are a number of sites (at random positions on the chain) with strength (distributed randomly within an interval) added to the on-site or hopping potential during the control process whereas the target state is still the edge state of the original chain without fluctuations. We should confirm that the positions and strength of these fluctuations are assumed fixed during the control process even they are random. We examine numerically the robustness of both control schemes for these two kinds of fluctuations by fidelity at time $t$=3000 vs. the number and strength of these fluctuations in Fig. \ref{SDEOH} for these two kinds of control schemes. From this figure, it can be seen that, with the increasing of the number and strength of fluctuations, fidelity would decline. This indicates the hindering effect of these fluctuations to the preparing procedures. However these fluctuations would change the original energy spectrum since they destroy the structure of the chain. Thus too many or fluctuation with too large amplitude would destroy this model which needs further discussion. Whereas in a range of the number and strength of these fluctuations, the fidelity can reach a high value which manifests both control schemes are robust against both kinds of fluctuations.

\begin{figure}
	\includegraphics[width=7.7cm]{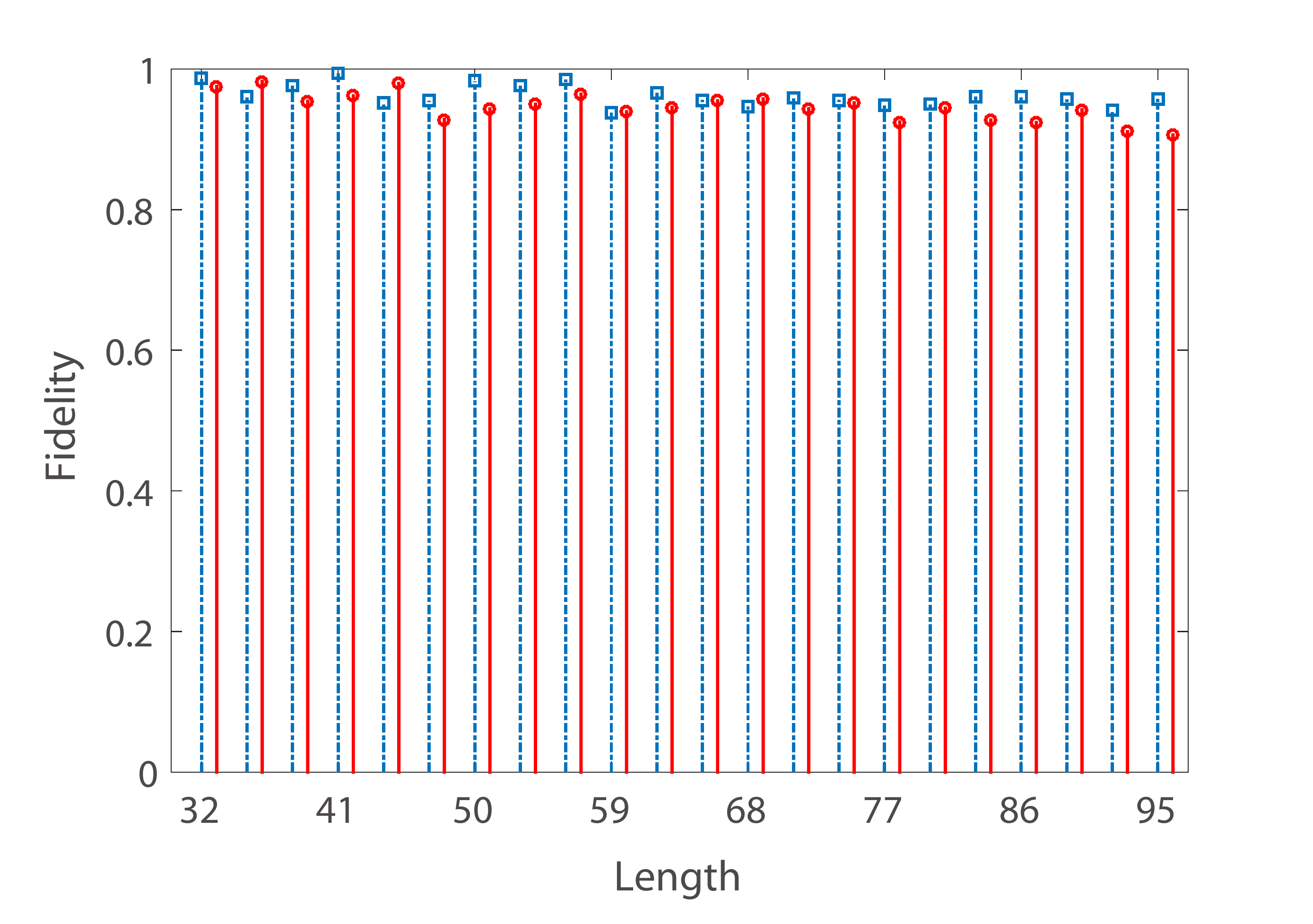}
	\caption{Fidelity at time $t$=3000 vs. several lengths of  the
		chains with the configuration 3$m+d$ while $d$=2, where the other
		parameters are same to those in Fig. \ref{FDE}. The nail graph with
		square heads denote the control results for Lyapunov-A scheme while
		those with circle heads denote the results for Lyapunov-B scheme.}
	\label{length}
\end{figure}

Since chains with different lengths may be the controlled objects, to examine the expandability of both control schemes, we apply them to such 3$m+d$ chains with $d$=2, within the length of $\langle32,35,...,95\rangle$ sites. `3' here reflects the periodic character of the chain, `m' is a integer and `d' is the remainder of the length of the chain divided by 3. In this length scope, the fidelity can reach a high value at time $t$=3000 even with slightly decrease with lengthening of the chain which is shown in Fig. \ref{length}.

\begin{figure}
	\includegraphics[width=7.7cm]{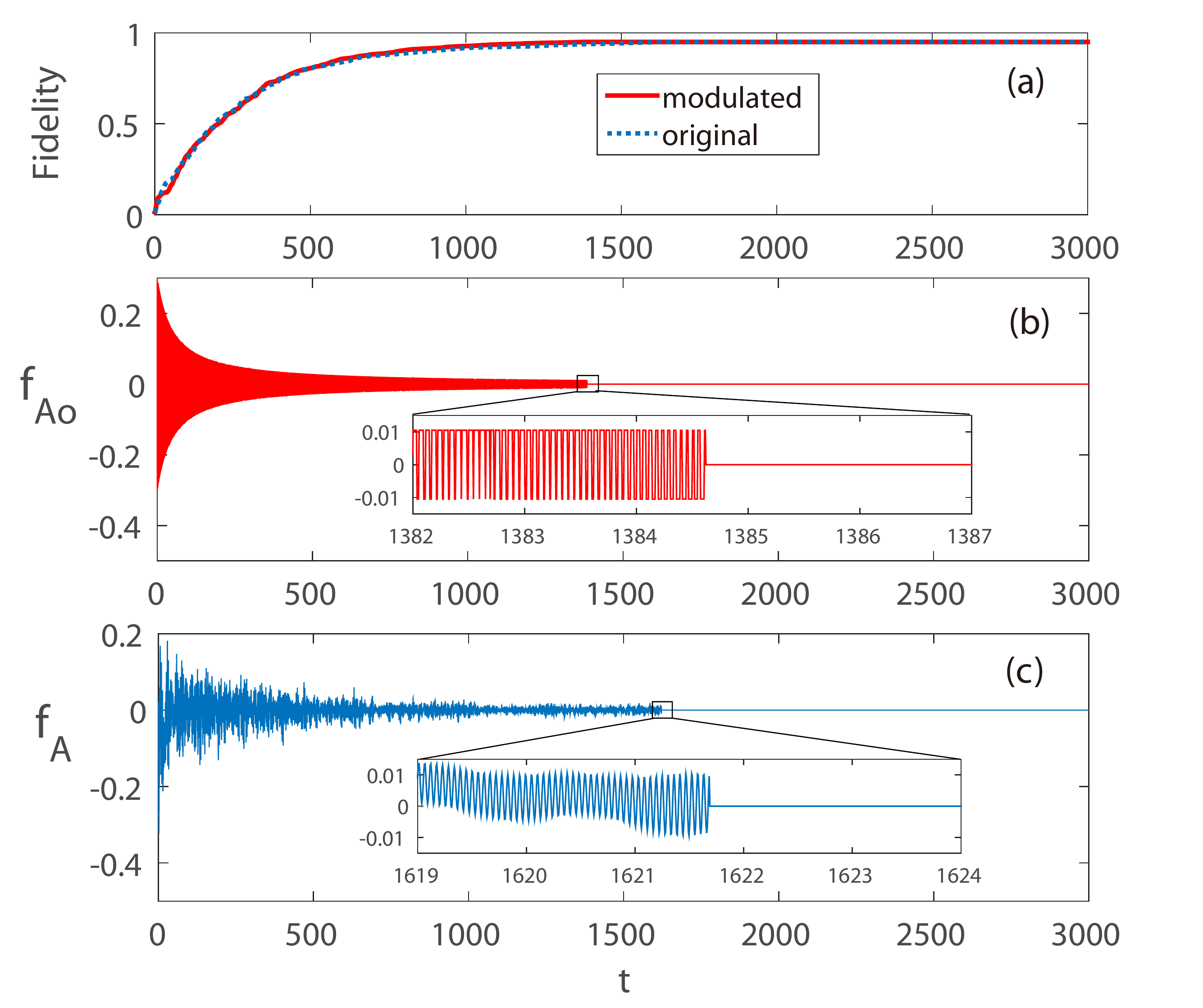}
	\caption{Compare of the results for amplitude-modulated   wave
		control and the original Lyapunov-A scheme. $f_{Ao}$ denotes the
		amplitude-modulated control field in Lyapunov-A scheme, while
		$f_{A}$ is the unmodulated one. We have set $A=0.3$ and $\kappa=50$
		in (\ref{mbb}). The control fields are shut down when the fidelity
		reach 0.95. The other parameters are same to those in Fig.
		\ref{FDE}. } \label{od}
\end{figure}

\begin{figure}
	\includegraphics[width=7.7cm]{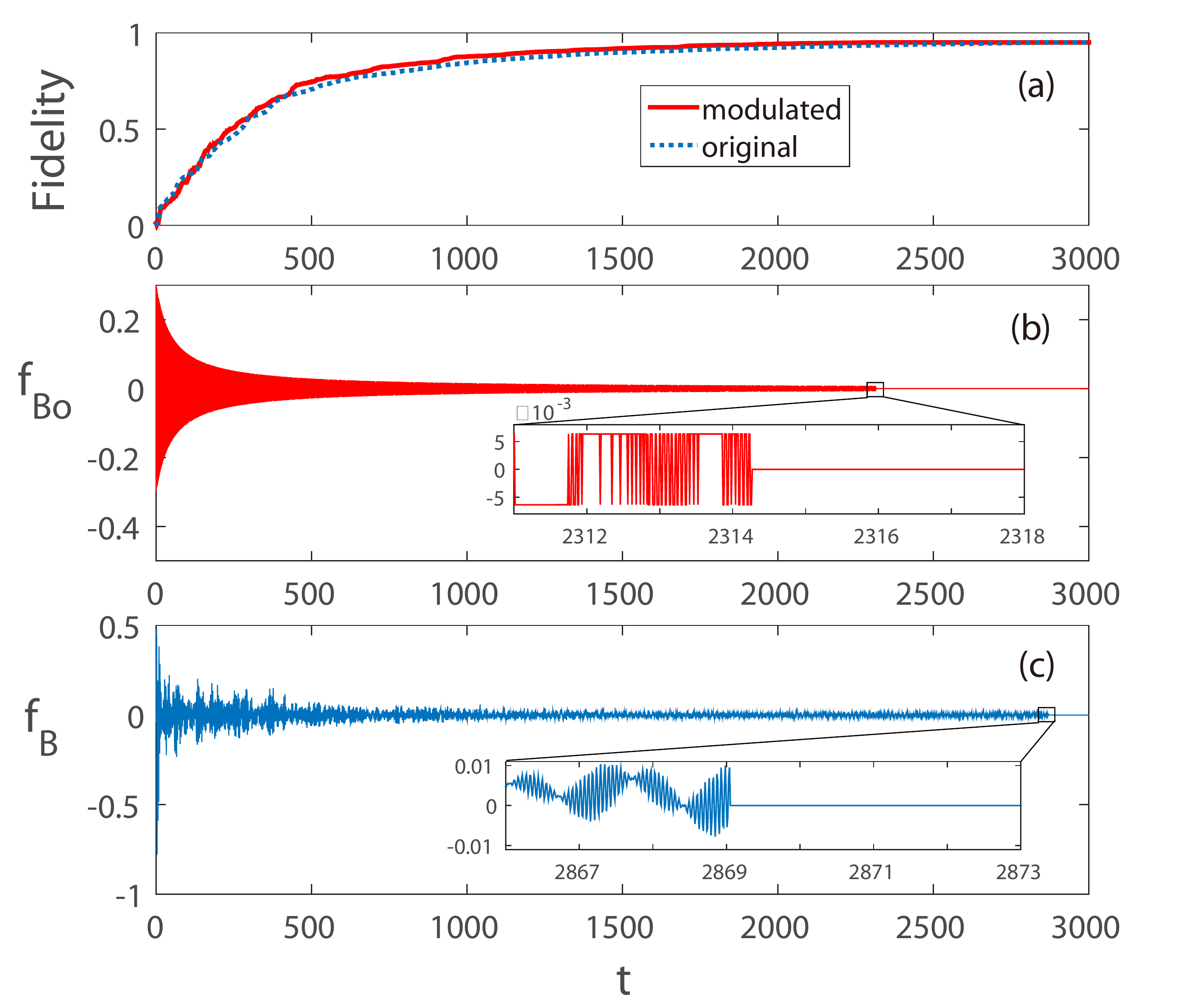}
	\caption{Compare of the results for amplitude-modulated   wave
		control and the original Lyapunov-B scheme.  $f_{Bo}$ denotes the
		amplitude-modulated control field in Lyapunov-B scheme, while
		$f_{B}$ is the unmodulated one. We have also set $A=0.3$ and
		$\kappa=50$ in (\ref{mbb}). The control fields are shut down when
		the fidelity reach 0.95. The other parameters are same to those in
		Fig. \ref{FDE}.} \label{oe}
\end{figure}

Compared to the original designed control fields such as those in Fig. \ref{FDE}, more regular control fields may be desirable in practice since regular control fields are easier to be generated than the original one. To achieve such a control field, we inspect the property of Lyapunov control. In this control, the sign plays a more important role than the amplitude of the control field since the sign of the control field determines the decreasing trend of the positive Lyapunov function but the amplitude determines the decreasing rate. An amplitude-invariant square waves has been used to replace the unmodulated control field \cite{PRA86022321} to complete a control process. However when the fidelity approaches 1, the evolution of the controlled system would be more sensitive to the amplitude of control field, i.e. the sign of the control field would switch more frequently between `+' and `-' if its amplitude stays invariant. Indeed in general, control fields with time-dependent envelopes can be used to realize a control goal. We use

\begin{eqnarray}
f_c(t)=\left \{
\begin{array}{rl}
F(t),~~~ f_c(t)>0, \\
-F(t),~~~f_c(t)<0, \\
\end{array}\label{bangbang}
\right.
\end{eqnarray}
here $F(t)>0$ is the control field modulated by a time varying envelope. There are various kinds of envelope functions which can be employed to realize a control goal. Considering the sensitivity of the sign of the control field to its amplitude when the state approaches the goal, we choose an envelope for the control field in terms of time $t$ as
\begin{eqnarray}
F(t)=\frac{A}{1+\kappa t}, A,\kappa >0.\label{mbb}
\end{eqnarray}
By these modulated control fields, the fidelity can reach 0.95 at time less than the results in the unmodulated cases as is shown in Fig. \ref{od} and Fig. \ref{oe}. Here $A$ and $\kappa$ in (\ref{mbb}) can be tuned flexibly. Finally, even the control fields are obtained by closed-loop simulation, whereas in light of Lyapunov control strategy, control fields with the same profile may be used in open-loop control process to prepare the edge state.

\section{summary} \label{sum}
The fermionic chain discussed in this paper can be created by loading cold atoms in optical lattice. Such an optical lattice can be created by two standing waves formed by laser beams of different wavelengths. This chain can be mapped to a ring with the same periodic structure mathematically. And it has attractive spectrum, in which the number of isolated eigenenergy depends on the hopping and the modulated phase $\delta$ in the Hamiltonian. After specified a 3$m+d$ fermionic chain defined in the text, we present proposals to prepare
an edge state by quantum Lyapunov control with state distance (Lyapunov-A) and state error (Lyapunov-B) schemes. By both control schemes, an initial state with equal population on each site can be steered to the edge state with high fidelity in a wide range of control parameters. In the simulation, we have chosen 100 site-occupation random initial states to show the validity of the control. And both schemes are available in the presence of fluctuations in on-site energies and hopping potentials. To reduce the difficulty in realization, the control fields can be replaced with amplitude-modulated ones. This is because the sign plays a more crucial role than the amplitude of the control field to achieve a control goal. Such control methods provide ways to explore edge state for topological materials which possesses novel properties.

\section*{ACKNOWLEDGMENTS}
This work is supported by the National Natural Science Foundation of China (Grant No. 11534002 and 61475033).

\end{document}